\documentclass{article}

\usepackage{arxiv}

\usepackage[utf8]{inputenc} 
\usepackage[T1]{fontenc}    
\usepackage{hyperref}       
\usepackage{url}            
\usepackage{booktabs}       
\usepackage{amsfonts}       
\usepackage{nicefrac}       
\usepackage{microtype}      
\usepackage{lipsum}		
\usepackage{graphicx}
\usepackage{natbib}
\usepackage{doi}

\usepackage{amsmath}
\usepackage{amsfonts}
\usepackage{amssymb}

\title{EPOC Emotiv EEG Basics}

\author{ \href{https://orcid.org/0000-0001-8155-0124}{\includegraphics[scale=0.06]{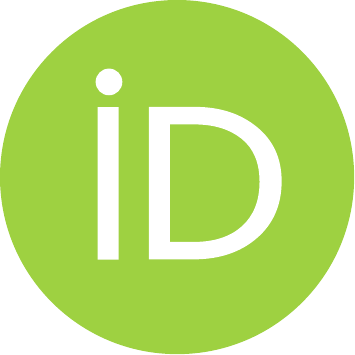}\hspace{1mm}Rodrigo Ramele} \\
	Department of Computer Engineering\\
	Instituto Tecnológico de Buenos Aires (ITBA)\\
	Buenos Aires, Argentina \\
	\texttt{rramele@itba.edu.ar} \\
	\And
	\href{https://orcid.org/0000-0000-0000-0000}{\includegraphics[scale=0.06]{orcid.pdf}\hspace{1mm}Ana Julia Villar} \\
	Department of Computer Engineering\\
	Instituto Tecnológico de Buenos Aires (ITBA)\\
	Buenos Aires, Argentina \\
	\texttt{ajvillar@itba.edu.ar} \\
	\And
	\href{https://orcid.org/0000-0000-0000-0000}{\includegraphics[scale=0.06]{orcid.pdf}\hspace{1mm}Juan Miguel Santos} \\
	Department of Computer Engineering\\
	Instituto Tecnológico de Buenos Aires (ITBA)\\
	Buenos Aires, Argentina \\
	\texttt{jsantos@itba.edu.ar} \\
}

\renewcommand{\headeright}{Technical Report}

\hypersetup{
pdftitle={Report: EPOC Emotiv EEG Basics},
pdfsubject={q-bio.NC, q-bio.QM},
pdfauthor={Rodrigo Ramele, Ana Julia Villar, Juan Miguel Santos},
pdfkeywords={EEG, Alpha Rythm, EPOC Emotiv},
}

\begin{document}
\maketitle

\begin{abstract}
	This document provides some basic guidance to start working with the EPOC Emotiv neuroheadset device and describes how to use it to perform basic Brain-Computer Interface (BCI) research.  A brief tutorial on how to set up the device, from its electrophysiological point of view, as well as a description and practical code to perform some basic analysis, is explained.  A basic experiment is introduced to detect one of the oldest and, indeed, quite still valuable electrophysiological correlate, visual occipital alpha waves, or Berger Rhythm.  An additional experiment is expounded where the power spectrum of alpha waves is reduced when a subject is affected by background cognitive disturbances.  This document also briefs about the extraction of information by using the EPOC Emotiv library and also with python Emokit package.   This report presents a basic guide on how to use EEGLAB + MATLAB, as well as python stack to perform the neurophysiological analysis.    Finally, a basic analysis on different feature extraction and classification methods is provided.  
\end{abstract}


\section{Introduction}
During the last couple of years, there has been an incredible advancement in technologies that allow to inspect human brainwaves in the form of Electroencephalographic signals, and use them as input to digital devices.  One of the earliest development in this trend has been the EPOC Emotiv device created by the Emotiv Inc.  This biopotentials measuring unit has been extensively used in BCI applications~\citep{c21}.  It provides an excellent balance between price and quality, with the additional benefit that it can be used wirelessly.

This work is a technical report, describing the most salient characteristics of this device and how to use them.  It is brief, intended to be read and used by anyone who wanted to start working with this device.  Additionally, it details how to use the standard APIs and its usage from MATLAB (Mathworks Inc., Natick, MA, USA), and a basic analysis by using EEGLAB.  Moreover, it describes how to use Emokit, an open source version of the USB dongle driver, and how to read the signals from python. 

This work unfolds as follows. In Section~\ref{sec:setup} introduces the characteristics of the device and their electrical properties. The next part, Section~\ref{sec:wearing}, describes several details and tips on how to use the device properly, how to connect it to a host computer and how to clean it.  It moves towards the software side, and in Section \ref{sec:software} it explains the different available options to obtain the raw data from the device.  On Section~\ref{sec:methodology},  this report describes a basic experiment to detect visual occipital alpha waves and how to perform basic analysis of the signals using EEGLAB, which is complemented on the next Section~\ref{sec:python} with details about how to do it from python.  Finally, discussion, conclusions and remarks are described in the last sections.

\section{EPOC Emotiv Neuroheadset}
\label{sec:headings}

\begin{table}
	\caption{Datasheet}
	\centering
	\begin{tabular}{ll}
		\toprule
		\cmidrule(r){1-2}
		Characteristic     & Value     \\
		\midrule
		Channels &  AF3, F7, F3, FC5, T7, P7, O1, O2, P8, T8, FC6, F4, F8, AF4 \\ 
		& (P3 and P4 can be additionally used but the signal is generally degraded because \\
		& they are used as reference in the standard reference configuration)   \\
		Sampling  & Sequential, only one ADC \\
		Sampling Rate & 128 Hz (internally it is performed at 2 KHz, and later downsampled) \\
		Resolution & 16 bits (14 usable bits) 1 LSB  = 1.95 $\mu V$  \\
		Bandwidth & 0.2 - 45 Hz, Notch filters at 40 and 60 Hz \\
		Dynamic Range & 256m Vpp \\
		Coupling & AC \\
		Connectivity & Proprietary wireless protocol, 2.4 GHz \\
		Battery & Li-Po \\
		Battery Life & 12 hs \\
		Impedance Measurement & Proprietary patented protocol  \\
		\bottomrule
	\end{tabular}
	\label{tab:table}
\end{table}

\section{Setup}
\label{sec:setup}

\begin{figure}
	\centering
	\fbox{\includegraphics[scale=0.7]{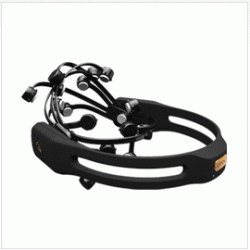}}
	\fbox{\includegraphics[scale=0.55]{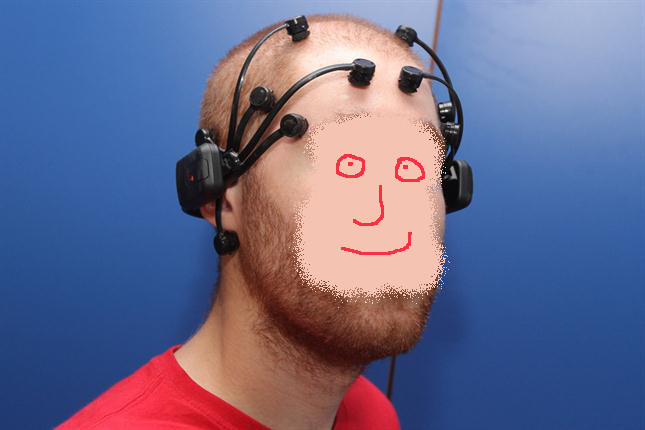}}
	\caption{The EPOC Emotiv headset device (version 1).  The device contains 16 electrode positions to provide a good-enough 14 channel's EEG signal. The device fits like a headset, and each electrode fits into their aimed position by a plastic and flexible lever. Hairless subjects are easier to work with.}
	\label{fig:device}
\end{figure}

The device can be seen on Figure~\ref{fig:device}.  It contains 16 plastic electrode slots.  Each slot is a plastic cylinder which contains a golden electrode plate at the bottom of it.  Small pads dumped with saline solution can be inserted on small plastic cases that can be screwed on each electrode slot.  These small plastic cases have a golden plate at the bottom, that make contact once they are fully screwed into each slot. These plastic cases can be bought separately and the entire set of electrodes can be replaced.
		
The device comes additionally with a USB dongle which works with a proprietary UHF communication protocol.   When this dongle is plugged in and the device is turned on, it will be ready to transmit all the data packages directly to the host computer.

\subsection{Important Notice 1}

The USB dongle works much better when it is not physically attached directly to a USB port on the host computer.  It works better when it is attached to a USB extension cable, and located around 50 cm from the computer.  Otherwise, some HDMI cables, for instance, that may be also attached to the computer can interfere with the dongle and the connection is easily lost.

\subsection{Important Notice 2}
Although the documentation states that the electrode pads should be lightly damped in saline solution, as more damped they are, the connection is actually better.   Naturally, this implies that the corrosion on the golden plate of each electrode plastic accumulates faster.

\subsection{Important Notice 3}
The electrode pads are stored in a 14-slots  small plastic case.  This box contains a bigger pad that can be also damped in saline solution and this really helps to have the electrode pads wet enough for the next recording season.  Otherwise, they get dry and turn very rigid and it is much more difficult to damp them again to work properly the next time.

\begin{figure}
	\centering
	\fbox{\includegraphics[scale=0.1]{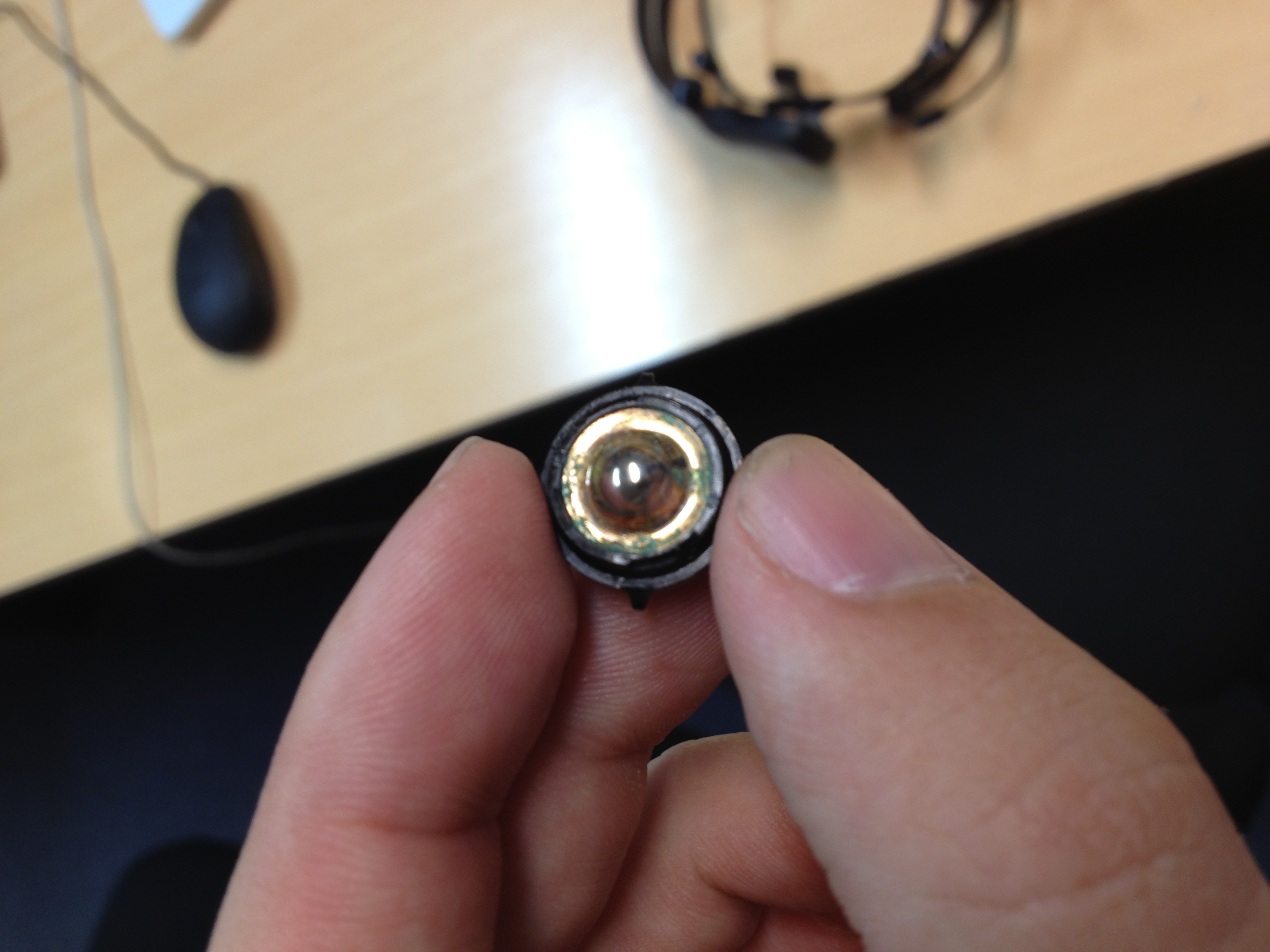}}
	\caption{Back view of the electrode plastic cases with the golden plate filled with corrosion.  Funny as it may seem, as corrosion builds up, the signal improves until it crosses a threshold where the signal degrades which mostly coincides when the moment when the corrosion is so high that the plate detaches from the plastic case.}
	\label{fig:electrode}
\end{figure}

\section{Wearing the device}
\label{sec:wearing}

The pads must be first separated from their plastic cases.  After damped them with enough saline solution, they can be inserted into each electrode case, verifying  they fit deeply enough to touch the bottom of the electrode slot, where the golden plate electrode is located.  Each one of the electrodes can be screwed, very \textbf{very} lightly inside each one of the electrode cases of the device.  The mechanism to hold each case against the electrode slot depends on a very small and fragile plastic lever that can be easily teared apart and if it breaks, will turn the electrode slot useless and it will have to be replaced.

The electrodes CMS and DRL (Figure~\ref{fig:emotivlocations}) must be adjusted against the mastoid (behind both ears) and T7 and T8 should be located just behind and just above the ears.  You should put your thumbs on CMS and DRL electrodes and your index fingers on T7 and T8, forming a sharp V letter with your fingers with ears in the middle.

Any EEG device requires a careful consideration of the electrophysiciological impedance~\citep{van2006signal}.  The Emotiv device has a very good system that provides this information from the API, and it provides a very easy and straightforward application to test it visually.
	
\subsection{Electrode adjustment}

Once the dongle is connected, a green light will start to blink.  At that time, the headset can be turned on, and put on the head of a participant subject.   It is then possible to open any program of preference (see Sections 2 and 3) to verify the impedance of each electrode.

This is by far, the most troublesome part of any electrophisiological procedure and it is the likely reason why graduate students don't want to perform experiments too often.  It can take around 20 minutes to properly set up all the electrodes in a painful iterative step-by-step procedure:

\begin{itemize}
\item Embed on saline solution the pad on the electrode cylinder plastic cases.
\item Press firmly the electrode to the scalp, and adjust its position a little bit to allow the electrode arm level to exert a force on the scalp. 
\item Reset the electrode configuration by pressing at the same time with two fingers the CMS - T7 and DRL - T8 electrodes (they act like a reset switch when pressed).
\end{itemize} 

This procedure must be repeated until all the electrodes are green (bellow $5 k \Omega$).  

\subsubsection{Emotiv Research Edition SDK}

This guide explains the procedure to use the Emotiv Research Edition SDK v1.0.0.5.  After installation of the program  on a windows computer, the program \textit{TestBench} can be used to measure the impedance of each electrode.  The nominal and reference value should be bellow  $5 k\Omega$.  Using TestBench this is represented by having all the electrode positions (Fig ~\ref{fig:emotivlocations}) in green.  A value of yellow represents a-not-so-perfect but still valid impedance while the color red means that the signal quality is almost unusable.  Black means that the electrode is electrophysiologically disconnected.

\subsection{Saline Solution}
Regular saline solution is required to embed all the small black electrode pads, to allow a good connection with the scalp.  As stated by the manufacturer, there are no special requirements on the type of solution:  "the required saline solution, contains a saline content of $0.5\%$ - $4\%$, better around $0.5\%$ to avoid electrodes corrosion;  non-allergic solution fitted with anti-microbials (you can do it yourself by adding $4\%$ isopropyl alcohol).  Anyway, multi purpose contact lens solution will work perfectly".

\subsection{Connection}

The connection takes time to establish, but once it works, it does for very long periods. The battery is excellent, and the device can run for at least enough time that the wearer won't like to use it anymore (i.e. the battery will not be the problem).

\subsection{Cleaning}

The golden plate on each electrode slot arm must be cleaned with a match wrapped in dry cotton, very lightly.  There is a little bit of corrosion that accumulates but if taken care meticously it should be very faint.   The electrode plastic case can be cleaned with a solution of water and soap.  The big problem that arises if you try to keep these electrodes clean, is that the connection quality drops, and it does that significantly.   The green material that accumulates is conductor and the best connection is actually obtained when there is some of it building up on the electrode (see Figure~\ref{fig:electrode}).

\section{Using MATLAB and EEGLAB on Windows}
\label{sec:software}
\subsection{Software}

\begin{itemize}
	\item Emotiv Research Edition
	\item Matlab 2011 R2
	\item EEGLAB
\end{itemize}

The official driver from Emotiv, comes with a sample program that describes how to retrieve the signals from the device and use them from a MATLAB program.  To do so, it is required to have the 32-bit version of MATLAB and the original DLL libraries for Windows.  At the same time, MATLAB needs to be configured to be able to run MEX files.  In brief, it is a 3-step process:  install a compiler for C (e.g. Visual Studio 2008), configure MEX compiling from MATLAB with \texttt{mex -setup} and finally include inside MATLAB's path, the directory where the Emotiv's dll can be found.

\subsection{From your head to a MATLAB array}

The file EDK.dll export the method \texttt{EE\_DataGet} which can be used to retrieve one full sample from the device.  This can be used to create the sample $\times$ channels EEG matrix.   The device sends all the 16 channels, plus additional channels with two gyroscope information and additional channels that can be used to verify the signal quality.

\begin{verbatim}
	if (readytocollect) 
    	   calllib('edk','EE_DataUpdateHandle', 0, hData);
    	   nSamples = libpointer('uint32Ptr',0);
        calllib('edk','EE_DataGetNumberOfSample',hData,nSamples);
        nSamplesTaken = get(nSamples,'value') ;
        if (nSamplesTaken ~= 0)
        	     data = libpointer('doublePtr',zeros(1,nSamplesTaken));
            	 for i = 1:length(fieldnames(enuminfo.EE_DataChannels_enum))
                	  calllib('edk','EE_DataGet',hData, DataChannels.([DataChannelsNames{i}]), 
                			data, uint32(nSamplesTaken));
                    data_value = get(data,'value');                                      
                    output_matrix(cnt+1:cnt+length(data_value),i) = data_value;                    
                end	                
                nS(cnt+1) = nSamplesTaken;
                cnt = cnt + length(data_value);
        end
	end
\end{verbatim}

\begin{figure}
	\centering
	\fbox{\includegraphics[scale=0.28]{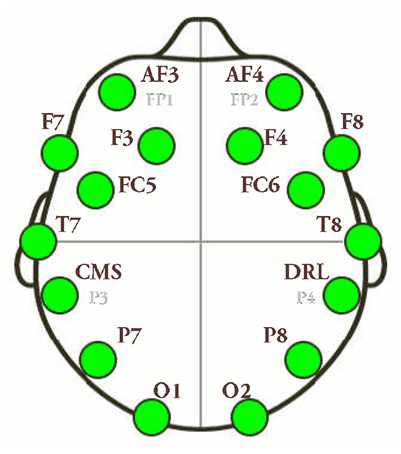}}
    \fbox{\includegraphics[scale=0.5]{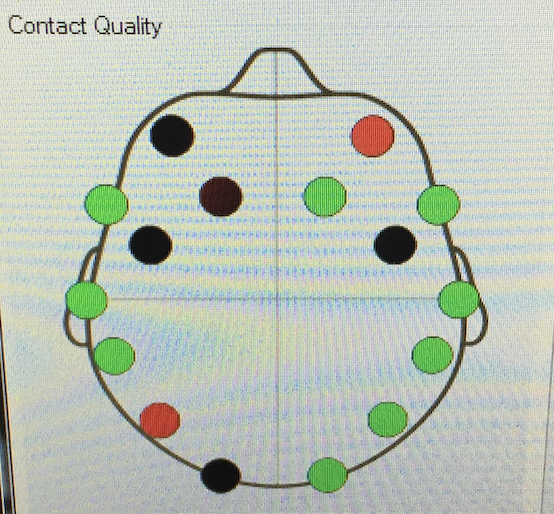}}
	\fbox{\includegraphics[scale=0.3]{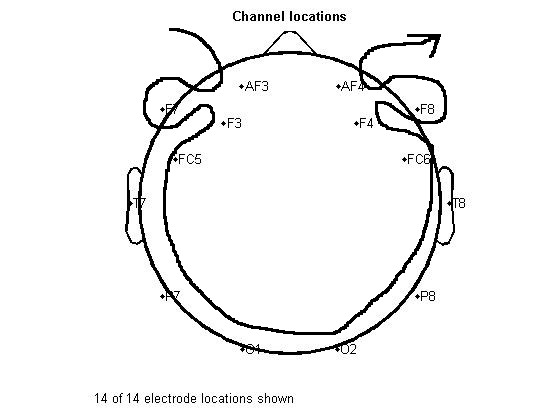}}
	\caption{Channel data position inside the received EEG Data Frame and approximate location of each electrode according to the 10-20 international system.  It starts in AF3 and ends in AF4}
	\label{fig:emotivlocations}
\end{figure}

Inside the data stream received by this procedure on \texttt{output\_matrix}, values indexes from 4 up to 17 (MATLAB indexes start from 1).  The order of each channel is described in Figure~\ref{fig:emotivlocations}.

Each channel contains the value in $\mu V$ obtained from each electrode.  Keep in mind that, there is a constant drift of these values so the signal mean will not be kept constant, and will deviate, drifting and oscillating.  Hence, this drift (i.e. where the zero is located) should be subtracted from the signal, implementing a basic baseline removal procedure.  In MATLAB, this can be done by doing:

\begin{verbatim}
[n,m] = size(output);
output = output - ones(n,1) * mean(output,1);
\end{verbatim}

As you can see, the MATLAB code to extract the EEG signals is basically a wrapper to the EMOTIV Win32 libraries.  Hence, you can remove MATLAB from the equation, and access the raw information directly from C++.  This is particularly useful if your intention is to implement any kind of real-time processing of the signals, because in that case MATLAB is a burden that you will not be willing to afford.  The code to do it is straightforward and can be obtained from the following repository: \url{https://github.com/faturita/epoceeg}~\citep{EPOCEEG}.

\subsubsection{EEGLAB}
This acclaimed MATLAB package is excellent to analyze EEG data, including the one produced by this consumer grade device.  It can be downloaded from \url{http://sccn.ucsd.edu/eeglab/} into a local folder and the only required thing to use it is to add that folder to MATLAB's path~\citep{Delorme2004}.

To be able to use this package, it helps to have the custom specification of channel localizations.  The following structure can be used to create a txt file \texttt{epoc.elp} containing this information:

\begin{verbatim}
346
     EEG	     AF3	 -75.803	 -67.539	      85
     EEG	      F7	 -95.055	 -36.087	      85
     EEG	      F3	 -62.027	 -50.053	      85
     EEG	     FC5	 -73.482	 -20.668	      85
     EEG	      T7	 -95.973	       0	      85
     EEG	      P7	 -95.055	  36.087	      85
     EEG	      O1	 -92.698	  72.074	      85
     EEG	      O2	  92.698	 -72.074	      85
     EEG	      P8	  95.052	 -36.133	      85
     EEG	      T8	  95.973	       0	      85
     EEG	     FC6	  73.482	  20.668	      85
     EEG	      F4	   62.01	  50.103	      85
     EEG	      F8	  95.052	  36.133	      85
     EEG	     AF4	  75.803	  67.539	      85
\end{verbatim}

EEGLAB expects the EEG matrix to be in the form channels x samples, so it must be transposed from the one generated from Emotiv, and it can be used to create the EEGLAB structures required for further analysis:

\begin{verbatim}
[ALLEEG EEG CURRENTSET ALLCOM] = eeglab;
EEG = pop_importdata('dataformat','array','nbchan',14,'data','f3','srate',128,'pnts',0,'xmin',0);
[ALLEEG EEG CURRENTSET] = pop_newset(ALLEEG, EEG, 0,'setname','relaxingeyesclosed','gui','off');
EEG=pop_chanedit(EEG, 'load',{'epoc.elp' 'filetype' 'autodetect'});
[ALLEEG EEG] = eeg_store(ALLEEG, EEG, CURRENTSET);
EEG = eeg_checkset( EEG );
\end{verbatim}

\section{Methodology: Experimentation Protocol}
\label{sec:methodology}

The \textbf{Experiment One} is a basic procedure in EEG processing.  A subject is comfortably sitting looking towards a white wall, without any extra stimulus, and wearing an EPOC Emotiv Neuroheadset with 14 electrodes successfully connected, obtaining 5/5 on impedance measurement on every channel.   The subject is instructed to:

\begin{itemize}
\item Stay awake and relaxed for 100 periods of 1 seconds.
\item Stay awake but close their eyes for 100 periods of 1 seconds.
\end{itemize}

While performing the experiment, the subject was instructed to take care in avoiding sudden movements of the head, tongue, neck and avoiding  as much as possible blinking during the whole experiment.  These movements generate typical signal artifacts that are routinely encountered in EEG analysis, and they tend to generate low frequency but very strong signal components.  While closing their eyes, it is important that the subject do not close them after the experiment has started.  That could potentially generate a strong artifact generated by their eyes movements.

This same protocol is repeated with the addition of a loud unintelligible conversation in the background, which construed the \textbf{Experiment Two}.

\subsection{Results}

\paragraph{Experiment One}

\begin{figure}
	\centering
	\fbox{
	\includegraphics[width=8cm]{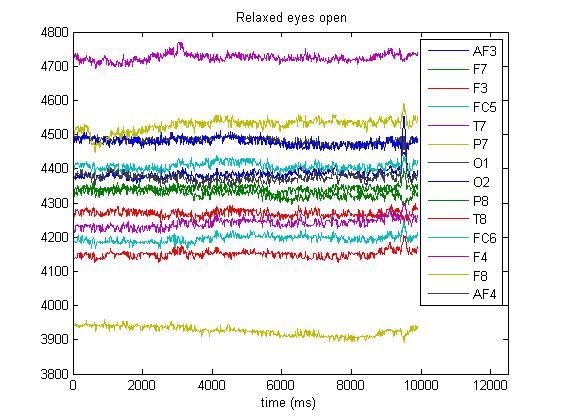}
	\includegraphics[width=8cm]{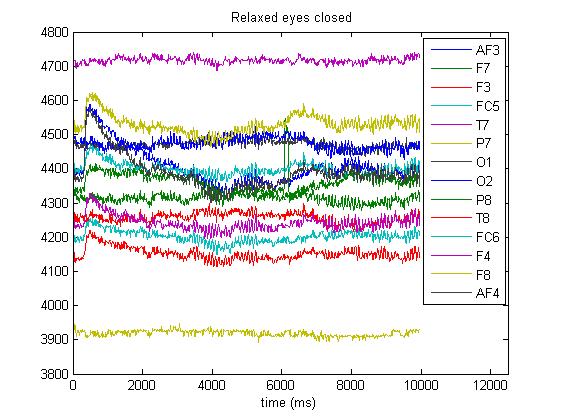}
	}
	\caption{Ten seconds of EEG signals corresponding to 14 channels,  (Left) Signals obtained while the subject seated relaxed with their eyes open and, (Right) with their eyes closed.  }
	\label{fig:plots}
\end{figure}

\paragraph{What kind of information can be obtained from the plots of these signals?}

Figure~\ref{fig:plots} shows a $10s$ segment of the obtained EEG signals for the 14 channels, for both conditions, open eyes and closed eyes.

\begin{itemize}
\item During the procedure with their eyes open, an artifact  present around $656$ ms, very subtle negative, particularly on T7.
\item For the same procedure, there is another artifact around $9300$ ms, which is positive on left channels, and negative on right channels.  This is very typical of ocular artifacts~\citep{Jiang2019}.
\item On the right figure, there is a very strong artifact at the beginning of the procedure, which is likely a muscular artifact~\citep{Dimitriadis2018}.
\item There is a sudden electrode disconnection on P8 around $6000$ ms. When the drift of the signal, a slow frequency wave, gets a sudden discontinuation, that is likely a disconnection event or some problem with the electrode impedance.
\end{itemize}

Additionally, from this visual inspection, it can be seen that there is a clear oscillatory component on the signals obtained when the subject kept their eyes closed (Figure~\ref{fig:plots} right).  This can be effectively analyzed in the frequency spectrum by using the Signal Processing package contained in MATLAB.  The following code could be used to obtain the frequency spectrum for the entire length of the obtained signal, for one single channel.

\begin{figure}
	\centering
	\fbox{
	\includegraphics[width=8cm]{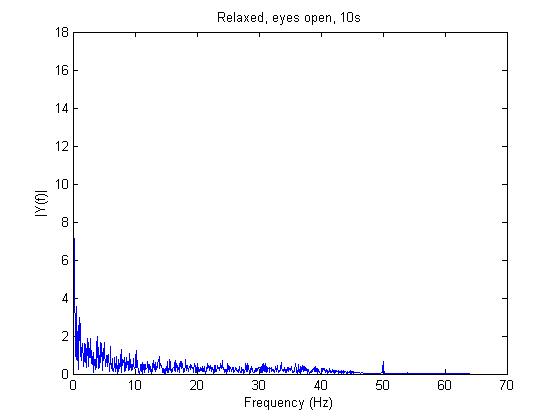}
	\includegraphics[width=8cm]{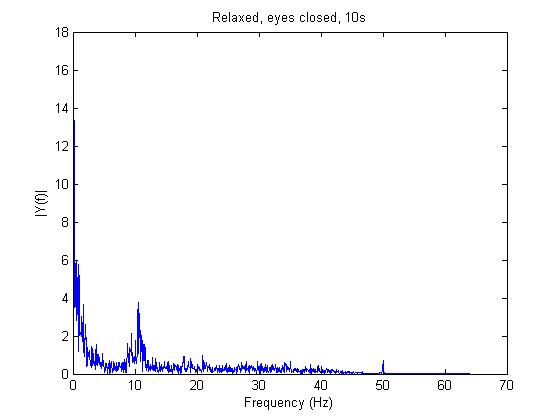}
	}
	\caption{Spectral representation of the signals segments from Figure~\ref{fig:plots}  for channel F3 of the obtained EEG signal, for both conditions, eyes open (Left) and eyes closed (Right).  The $10$ Hz peak, clearly visible when the eyes are closed, is due to the presence of the Berger Rhythm which is manifested by the presence of alpha waves. }
	\label{fig:spectral}
\end{figure}

\begin{verbatim}
Fs = 128;					# EPOC Sampling Frequency
T = 1/Fs;					# Sample time
L = size(signal,2);		# Length of the signal
t = (0:L-1)*T;				# Scaled Time Vector

% Add a pure 50 Hz control signal
x1 = 0.7*sin(2*pi*50*t) + signal;

NFFT = 2^nextpow2(L);# Approximate to the nearest power of two for efficiency.
Y = fft(x1, NFFT)/L;
f = Fs/2*linspace(0,1,NFFT/2+1);

% Plot single-sided amplitude spectrum
plot(f,2*abs(Y(1:NFFT/2+1)));
xlabel('Frequency (Hz)');
ylabel('|Y(f)|');
\end{verbatim}

This code generates Figures~\ref{fig:spectral} which correspond to the spectral information on the F3 channel for the timespan of the $10$ s segments plotted on Figures~\ref{fig:plots} for both conditions, eyes open (Left), eyes closed (Right).  You can see that there is an spurious $50 Hz$ signal added to the signal, that can be used to control that everything is in its place (i.e. this signal should be appear on the spectral figure).

\begin{figure}
	\centering
	\fbox{
	\includegraphics[width=8cm]{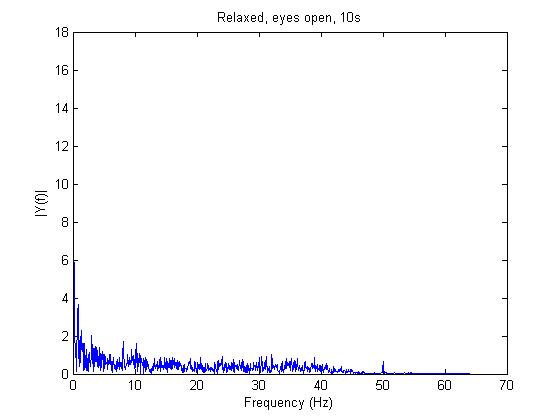}
	\includegraphics[width=8cm]{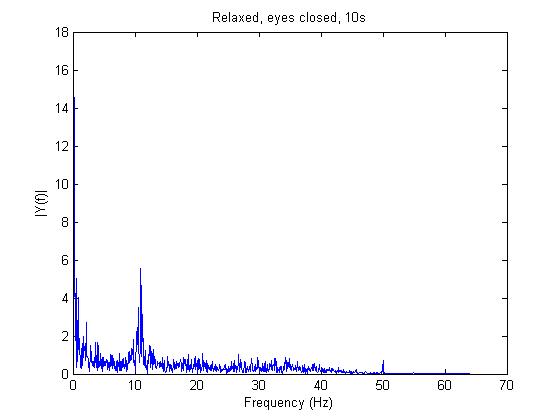}
	}
	\caption{Spectral figures for channel O2 of the obtained EEG signal, for both conditions, eyes open (Left) and eyes closed (Right).  The $10$ Hz peak, the hallmark of alpha waves, can be spotted on the right figure. }
	\label{fig:spectral2}
\end{figure}

From these figures, it can be spotted very clearly that there is a clear peak around $10$ Hz.  As these are spectral figures, this means that there is an oscilatory component of around 10 crests per second, which is more present when the subject closed their eyes.  These are called Alpha Waves~\cite{Sanei2007}, which were the first signals ever spotted from Electroencephalography.  They are actually characterized as 10  Hz, or more broadly between 8-12 Hz range.  They are physiologically consistent across subjects, though it has been reported inter- and intra- variations with functional cognitive implications~\citep{Haegens2014}. Moreover, they are associated with synchronous inhibitory processes and attention shifting.  These waves are also called Prominent Posterior Alpha or Posterior Dominant Rhythm~\citep{Schomer2010}.  They can appear on every region of the scalp but they are more prominent on occipital channels.  This pattern regularly appears when a person has their eyes closed, and it is hypothesized  that this is due an exhausting searching activity that analyze the visual cortex, while trying to find the absent visual stimulus~\cite{Sanei2007}.  This is clearly  shown, when we do the same spectral analysis but this time for occipital channels O8.  On Figure~\ref{fig:spectral2}, it can be seen that the accumulated power of the alpha wave signal is bigger than the one previously obtained for F3.  This suggest that there is more power on the occipital region and that alpha waves oscillations are actually stronger there.

\begin{figure}
	\centering
	\fbox{
	\includegraphics[width=8cm]{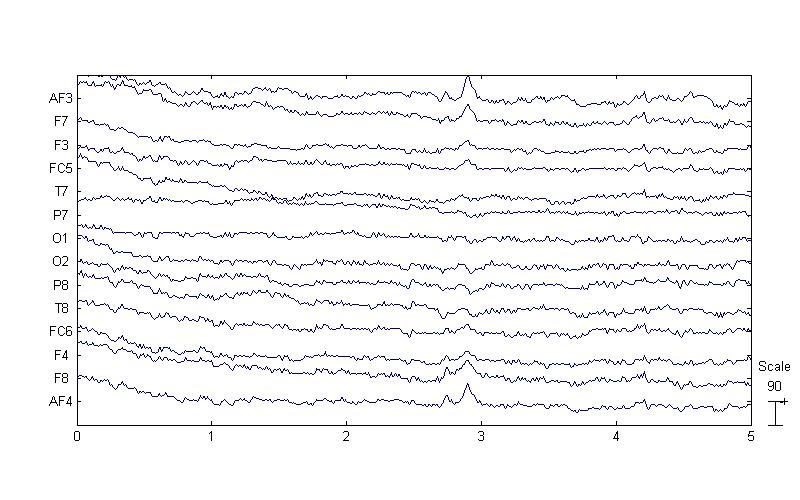}
	\includegraphics[width=8cm]{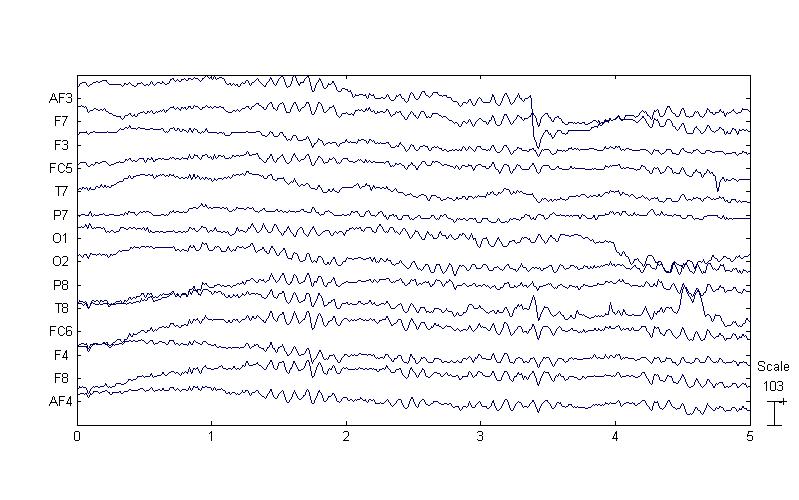}
	}
	\fbox{
	\includegraphics[width=8cm]{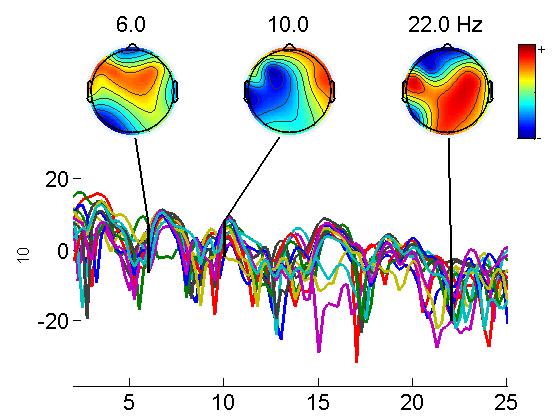}
	\includegraphics[width=8cm]{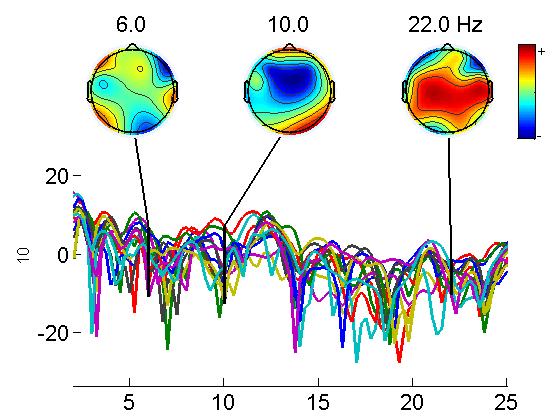}
	}
	\caption{Top: EEG signals corresponding to the 14 channels provided by EEGLAB, preprocessed and bandpass filtered.  Bottom: Scalp maps with forward projections that represent the contributions to the obtained spectrum on each frequency from each different brain region.  The presence of strong alpha waves (10 Hz components) on the occipital area can be seen on the figure from the Right.}
	\label{fig:proj}
\end{figure}

The package EEGLAB  can be also used to create scalp projections which can be seen on Figure~\ref{fig:proj}.  These are called \textbf{topomaps} and are handy to understand the power distribution in different frequencies across the scalps.  Top figures, describe $5$ s of signals for all the 14 channels, for both conditions.  EEGLAB applies a low pass filter on the signals, with a cut-off frequency of $20$ Hz.  Now, the alpha waves are more clearly visualized on the top-right figure, as small wiggles present on every channel (after the first second).  On the left figure, these wiggles are not present: this is generally term \textbf{alpha wave visual blocking with eyes open}.

From these diagrams it can be seen how 10 Hz components are more prominent on occipital areas while the subject is seated with their eyes closed, which may indicate a strong synchronization on the visual cortex due to the lack of stimulus.  On the topomaps at the bottom of Figure~\ref{fig:proj}, the topoplot of the right figure at 10 Hz, show more red on the occipital regions which is also what we encountered previously while we verified that the frequency component of occipital channels (O2) were stronger those of frontal channels (F3). 

\paragraph{Experiment Two}

Alpha waves are produced by a lack of stimulus and they represent cognitive attention shifting~\citep{VanGerven2009}. We may ask ourselves what would happen if there is a background audible conversation on the same room.  Could it be possible that this represents an alpha wave reduction while eyes are closed?  This question is what we address in this experiment.

\begin{figure}
	\centering
	\fbox{
	\includegraphics[width=8cm]{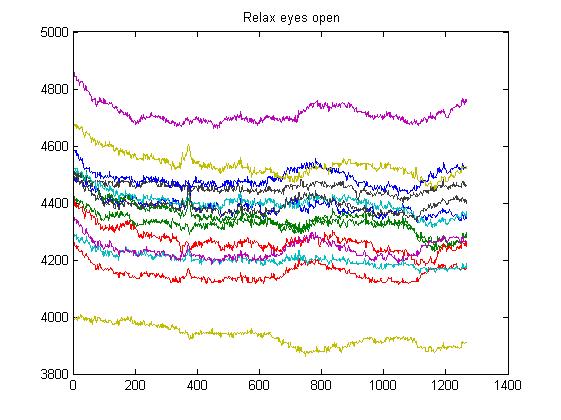}
	\includegraphics[width=8cm]{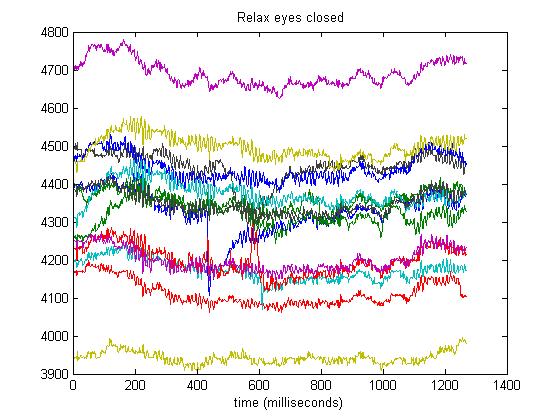}
	}
	\fbox{
	\includegraphics[width=8cm]{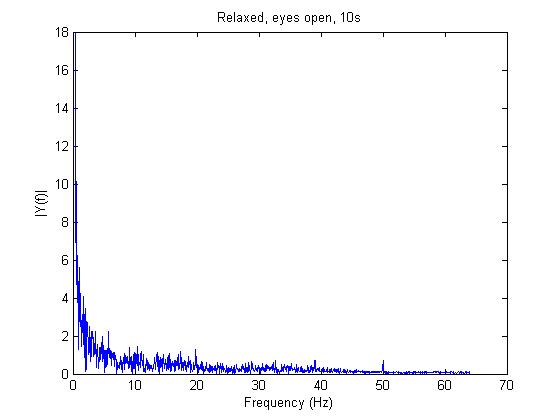}
	\includegraphics[width=8cm]{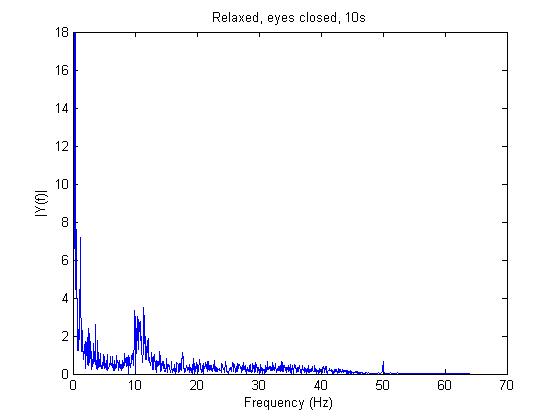}
	}
	\caption{Top: EEG signals corresponding to the 14 channels for both conditions on Experiment 2, where the closed eyes condition includes an audible conversation heard in the background.  Bottom: Spectrum figures of both conditions, eyes open (left), eyes closed with background hearing (right) on the channel O2.  Now the power of the $10$ Hz component is still present in relation to the conditions, but it is diminished in relation to previous experiment.}
	\label{fig:hearing}
\end{figure}

Figure~\ref{fig:hearing} shows the result of the same experiment, but on the eyes closed condition, a relative loud conversation is being heard in the background.  It is a conversation in a foreign language, unknown to the subject.  On the bottom-right subfigure, the 10 Hz peak is still present on the spectral chart though their power is comparatively diminished than those obtained for Figure~\ref{fig:spectral2}, and the frequency spectrum around 10 Hz is more distributed and less centered.  This is, of course, just a representative experiment, but these may suggest that there is now less space in the cognitive \textit{idle mode}, because some resources are now being used to the listening operation\footnote{This statement is totally speculative, naturally.}.

\section{Python processing}
\label{sec:python}

Python is now the undoubtedly Queen of Data Science, and with it it is also one of the Master of the Universe of Neuroscience processing tools~\citep{Gramfort2013,PythonNerv}.

\subsection{Emokit}

This section introduces us finally to a controversy that arose in Emotiv, and that can be easily extended to the boundaries of technological ecosystems.  This guide describes how to use the first EPOC+ 2013 Emotiv Research Device.  This version allowed access to the raw EEG signals, which is a key feature to encourage a thriving technological ecosystems of \textit{makers} that can create on top of the device.  However, later on, the company decided to severely restrict the access to these raw signals by a very expensive software license, to somehow control more strictly that ecosystem~\citep{Dadebayev2021}.

Despite of this, the software community found \textit{a way} to access the raw EEG information.  Emokit is a hack which allows to access the raw data from the EPOC Emotiv device.  Created by Kyle Machulis it is hosted now as a community project~\citep{Emokit}.   The version used in this guide is an updated fork at \url{https://github.com/faturita/python-nerv}.  The whole library is composed of just a single python script \url{https://github.com/faturita/python-nerv/blob/master/emotiv.py}.  This code uses HID to access the information from the Emotiv dongle, decrypts each frame, and access the raw data for all the sensors at the highest frequency possible.  

The library uses \texttt{gevent} python library.  

\begin{verbatim}
while True:
    KeepRunning = True
    headset = None
    while KeepRunning:
        try:
            headset = emotiv.Emotiv(display_output=False)
            gevent.spawn(headset.setup)
            g = gevent.spawn(process)
            gevent.sleep(0)

            gevent.joinall([g])
        except KeyboardInterrupt:
            headset.close()
            quit()
\end{verbatim}

And a function call that will be executed each time a new frame is available.

\begin{verbatim}
def process():
    ts = time.time()
    st = datetime.datetime.fromtimestamp(ts).strftime('%Y-%m-%d-%H-%M-%S')
    f = open('sensor.dat', 'w')
    readcounter=0
    iterations=0

    while headset.running:
        packet = headset.dequeue()
        interations=iterations+1
        if (packet != None):
            datapoint = [packet.O1, packet.O2]
            f.write( str(packet.gyro_x) + "\t" + str(packet.gyro_y) + "\n" )
            readcounter=readcounter+1

        if (readcounter==0 and iterations>50):
            headset.running = False
        gevent.sleep(0)
\end{verbatim}

On the \texttt{headset} object, besides all the available channels, there is also some extra information, particularly two gyroscope values: \texttt{GYRO\_X} and \texttt{GYRO\_Y}, which corresponds to headset yaw and pitch movements.

There is an additional field, called \texttt{COUNTER} which provides an incremental number which is assigned to each frame transmitted from the device.  This is very handy to try to determine if there is any frame that was lost, and apply some data missing technique to recover the values and maintain the same sample frequency.

\begin{figure}
	\centering
	\fbox{\includegraphics[scale=0.3]{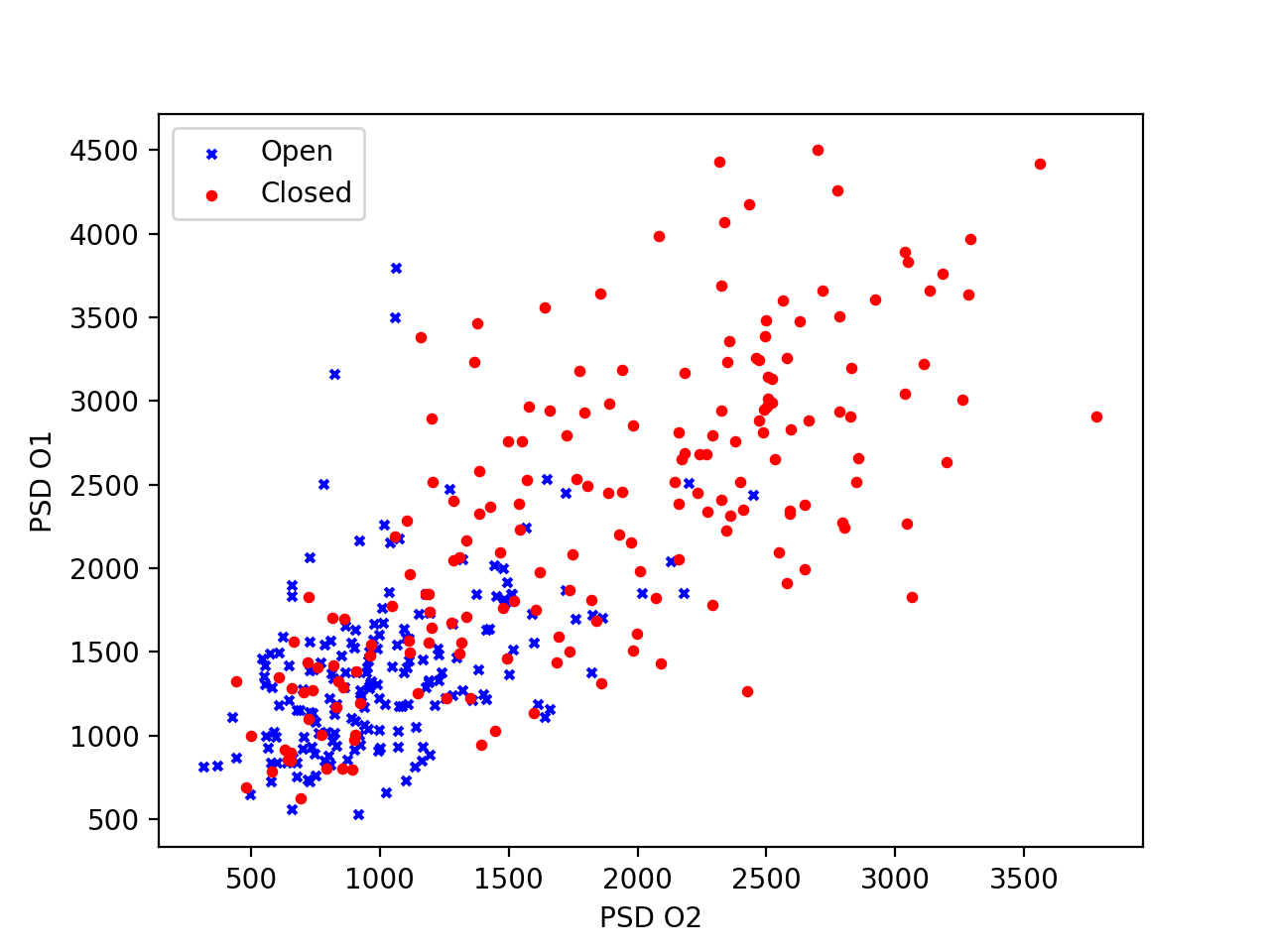}}
    \fbox{\includegraphics[scale=0.3]{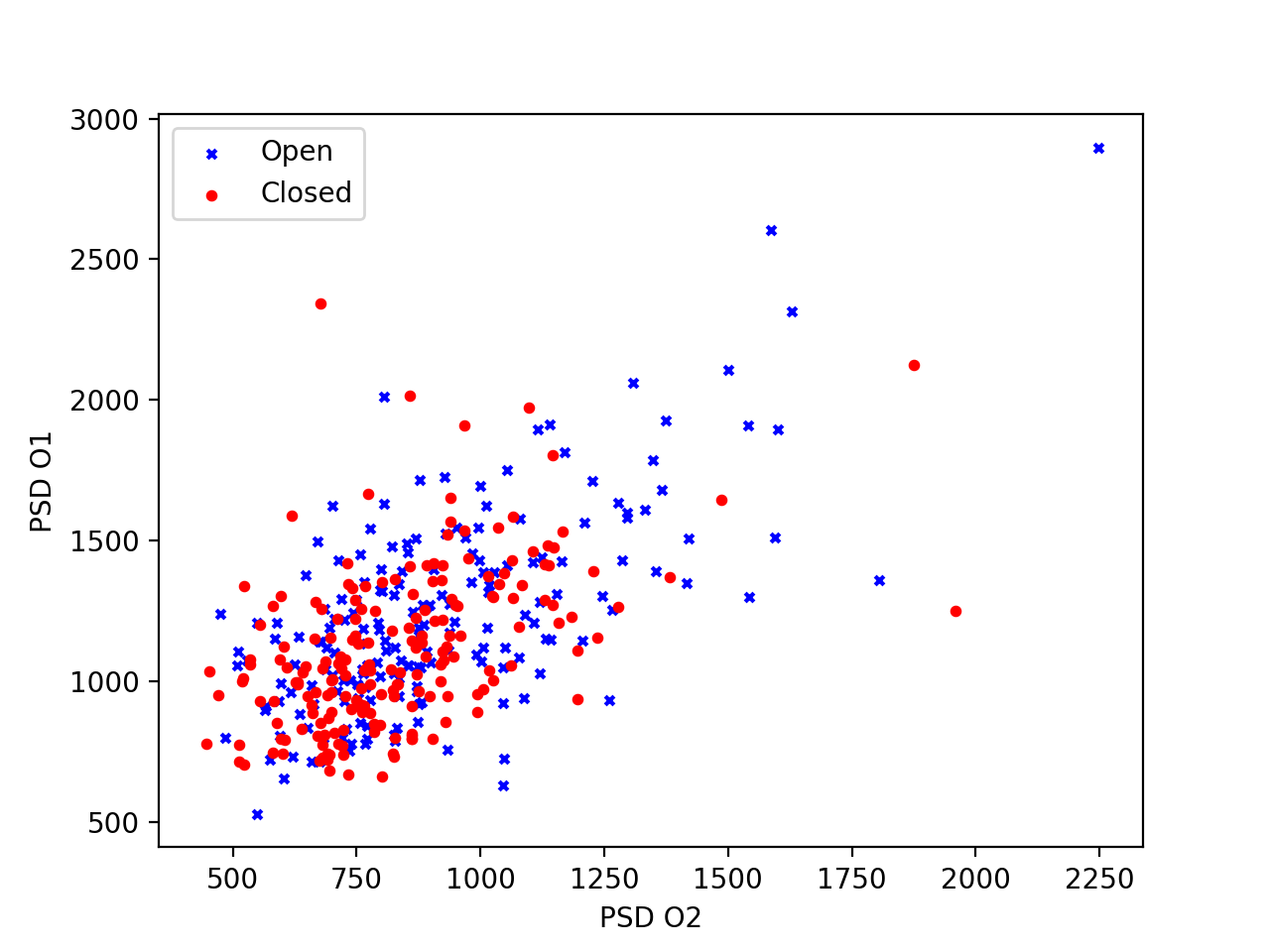}}
	\fbox{\includegraphics[scale=0.3]{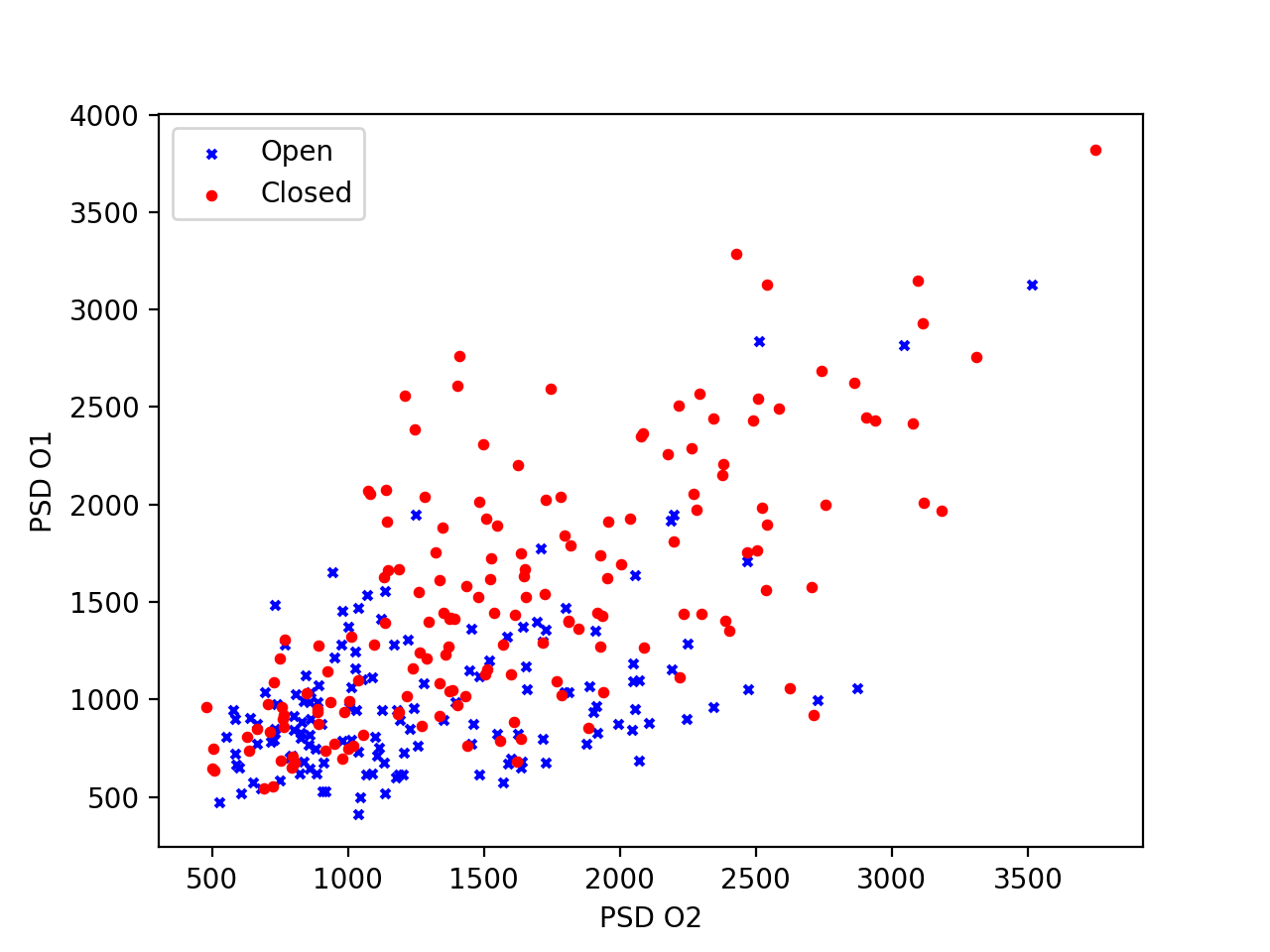}}
	\caption{Clustering of two dimensional features for Subjects 1, 2 and 3.  Features were obtained by applying a 1-second length sliding window with an overlapping of half window size, and calculating the PSD of channels O1 and O2.}
	\label{fig:clustering}
\end{figure}

\paragraph{Classification of the signals}

The experiment 1 was extended to two more participants.  The data for both conditions, is processed in python by implementing a sliding window with of 128 time points (which at 128 Hz is 1 seconds) with an overlapping of half size of the window.  From each window, a segment for O1 and O2 is extracted, and the Power Spectral Density (PSD) is calculated for each one of them.  This composes a 2-dimensional feature that is used to create the plots on Figure~\ref{fig:clustering}.

On these figures it can be seen that red points, corresponding to the eyes closed condition, are more concentrated in the upper right corner, which means more power on both channels at the same time.  Points corresponding to the first condition are more widespread.

It is important to note two of the subjects (1 and 3) there are clear differences on the distributions of the features, but their nature is different.  For subject number 2 the difference is more difficult to observe. The code that generates these figures are detailed here: \url{https://github.com/faturita/python-nerv/OfflineSignalAnalysis.py}.

Three basic classifiers were used using these features for each subject: Support Vector Machine with linear kernel~\cite{Rakotomamonjy2008}, Logistic Regression~\cite{Lotte2018} and finally a Multilayer Perceptron (MLP) of 2 hidden layers of 64 and 32 neurons each, with tanh activation functions for the two layers, and a sigmoid for determining the output value~\cite{Tjandrasa2018}.  Results can be seen on Table~\ref{tab:table}, where the obtained accuracy show the same results as in Figure~\ref{fig:clustering}, where the accuracy is significant for subjects 1 and 3, but around chance level for Subject 2.  

It is also important to remark that a simple MLP model using Keras~\cite{chollet2015keras} failed to obtain any result at all.  Granted, there are several hyperparameters that could be adjusted for the neural network but the SVM, and LDA classifiers were used in an out-of-the-box approach.  It is, perhaps, a reminder that neural networks are data cruncher techniques and for some bio-physiological applications data is very hard to obtain.  This is one of the limits of current Deep Learning approaches~\cite{Lotte2018}.

\begin{table}
	\caption{Classification accuracy values obtained for each subject, based on the features extracted for the data of experiment 1.}
	\centering
	\begin{tabular}{llll}
		\toprule
		Subject     & SVM  & LogReg  & MLP     \\
		\midrule
		Subject 1 & $\sim$75 \%    & $\sim$76 \%  & $\sim$49 \%   \\
		Subject 2 & $\sim$56 \%   & $\sim$55 \%  & $\sim$51 \%  \\
		Subject 3  & $\sim$70 \%   & $\sim$70 \%  & $\sim$58 \%  \\
		\bottomrule
	\end{tabular}
	\label{tab:table}
\end{table}

\section{Discussion}

This work is intended to be a validation procedure which helps to understand how to set up properly this device and at the same time verify a physiological marker that is the proof that obtained signals are meaningful.  Moreover, the current hype on BCI technology is extending this discipline to the tinkering, DIY and Maker movement where very pragmatic innovation always takes place.  Devices like the EPOC Emotiv have provided the means to deliver for this trend.

Future work could be conducted in terms of 
\begin{itemize}
\item Verify the impeadance of the electrodes to determine which is the right amount of saline solution
\item Test the alpha waves experiment in light depraving environments, instead of forcing participants to close their eyes.
\item Preprocess the signal effectively.
\item Study different types of signals, including the P300 which is a very interesting signal that is widely used in BCI.
\end{itemize}

\section{Conclusion}

TL;DR:  The device is excellent.  Although, the quality of the signal is not best, this is a real brainwave extraction device and you can peek what is happening inside the brain.  Additionally, all the issues while reading electrophisiological signals are present, so it is an extraordinary device for teaching and learning.  On the downside, Emotiv has chosen a negative path of heavy licensing with a monthly fee to the possibility of access raw EEG data. New devices manufactured by Emotiv contain a different protocol and Emokit does not work with them (yet :).

\bibliographystyle{unsrtnat}
\bibliography{references}  






\end{document}